# Tailoring of rhenium oxidation state in $ReO_x$ thin films during reactive HiPIMS deposition process and following annealing

M. Zubkins, A. Sarakovskis, E. Strods, L. Bikse, B. Polyakov, A. Kuzmin, V. Vibornijs, J. Purans

Institute of Solid State Physics, University of Latvia, Kengaraga 8, LV-1063, Riga, Latvia

## Abstract

Bulk rhenium trioxide ($ReO_3$) has an unusually high electrical conductivity and, being nanosized, has promising catalytic properties. However, the production of pure $ReO_3$ thin films is challenging due to the difficulty to stabilize rhenium in a 6+ oxidation state. Here we present a novel approach for the deposition of $ReO_x$ ($x \approx 1.6$–$2.9$) thin films using reactive high power impulse magnetron sputtering (r-HiPIMS) from a metallic rhenium target in a mixed $Ar/O_2$ atmosphere. The thin films were deposited in the gas-sustained self-sputtering regime, observed during r-HiPIMS process according to current waveforms. The influence of the substrate temperature, the oxygen-to-argon flow ratio and post-annealing at 250 °C in the air for 3 h on the properties of the films were studied. The as-deposited films have an X-ray amorphous structure ($a$-$ReO_x$) when deposited at room temperature while a nano-crystalline $\beta$-$ReO_2$ phase when deposited at elevated temperatures (150 or 250 °C). The amorphous $a$-$ReO_x$ can be converted into the crystalline $ReO_3$ with a lattice parameter of 3.75 Å upon annealing in the air. The surface morphology of the films is dense without detectable voids when elevated substrate temperatures are used. Various Re oxidation states are observed on the surface of the films in different ratios depending on the deposition parameters. All samples exhibit electrical resistivity on the order of $10^{-3}$ Ωcm and optical properties typical for thin metallic films.

## 1.Introduction

This study is dedicated to the deposition of rhenium oxide thin films by high power impulse magnetron sputtering (HiPIMS), and the possibility to change the oxidation state of rhenium ions upon thermal annealing in air. The chemistry of rhenium (Re) is characterised by a large number of oxidation states [1]. Rhenium oxides can contain cations with the oxidation states 4+, 6+, and 7+ with the corresponding crystalline phases $Re_2O_7$, $ReO_3$, and $ReO_2$, respectively. $Re^{7+}$ and $Re^{4+}$ are the most stable among them. Re oxides are also promising catalytic materials [2-5]. The most famous rhenium oxide is $ReO_3$ which is also called a “covalent metal”. $ReO_3$ is a red solid with an electrical conductivity comparable to metals due to the delocalised 5*d* electrons. It is also a negative thermal expansion material [6]. In addition, $ReO_3$ appears to be a suitable precursor to achieve nanosheets of $ReS_2$ on carbon fibres [7]. The oxide decomposes into $ReO_2$ and $Re_2O_7$ above 400 °C without melting. $ReO_3$ has perovskite-type structure ($ABX_3$) (space group *Pm-3m*) in which the larger cation (A) site in the centre of the cubic unit cell is empty [8, 9]. The structure consists of regular corner-sharing $ReO_6$ octahedra with the Re−O−Re angles equal to 180°. Materials with the $BX_3$ ($ReO_3$-type) structure have many attractive properties due to the connectivity and openness of the structure, for example, a negative thermal expansion coefficient [10], photocatalysis [11], thermoelectricity [12], and superconductivity [13]. There are several techniques for the synthesis of $ReO_3$ nanoparticles, which are summarized in Ref. [14]. Nanocrystalline $ReO_3$ thin films can be prepared by the annealing of amorphous $ReO_x$ films at approximately 250 °C in the air under an inert capping material [14]. Similarly, layered $WO_3$/

$ReO_3$ and $ReO_3/WO_3$ thin films have recently been prepared [15]. $ReO_2$ is a brown to black material that has two possible structural phases. A monoclinic $\alpha$-$ReO_2$ ($P2_1/c$) structure is similar to $MoO_2$ and $WO_2$ and forms below 300 °C. Orthorhombic $\beta$-$ReO_2$ ($Pbcn$) consisting of the corner and edge-sharing $ReO_6$ octahedra forms above 300 °C. In both phases, the distance between Re atoms is small, indicating a direct metal-metal interaction with the consequential high electrical conductivity. $Re_2O_7$ is yellowish, electrically insulating solid that melts at around 300 °C. The crystal structure ($P2_12_12_1$) consists of an equal number of almost regular $ReO_4$ tetrahedra and strongly distorted $ReO_6$ octahedra. The tetrahedra and the octahedra share corners through oxygen bridges and form a sheet-type structure with van der Waals bonds between the sheets. $Re_2O_7$ is volatile and easily converts into perrhenic acid ($HReO_4$) when it interacts with moisture [16]. Nonstoichiometric $ReO_3$ can be obtained by heating the complex of $HReO_4$ and dioxanes [17].

The physical properties of thin films are strongly influenced by the ion bombardment during the film growth in plasma-based deposition processes [18]. HiPIMS is the technique that provides the high ionised density fraction of sputtered species [19]. During the pulse, the plasma density grows up to approximately $10^{20}$ $m^{-3}$, which ensures high ionisation. However, the ionised flux fraction towards the substrate is almost always lower because the substrate is usually located outside the ionisation region and some of the ions in this region are attracted back to the target due to the negative potentials [20]. In addition, this effect also reduces the deposition rate (30–85% of direct current magnetron sputtering) [21]. There are several strategies to address these issues – modification of the magnetic field strength and topology [22], reduction of the sputtering pulse [23], use of chopped HiPIMS [24], or introduction of an additional ionisation zone further away from the target [25]. Nevertheless, the ionised flux can be controlled by an electric and magnetic field to adjust the properties of the growing film.

HiPIMS has been shown to provide dense and smooth films with good adhesion [21, 26], good control of the film's microstructure [27], improvements in mechanical [28] and other properties at low deposition temperatures [29, 30], and uniform coverage on non-flat substrates [31]. As HiPIMS is already a well-established coating technology in the field of hard coatings, it can become more relevant in the fields of catalytic or optoelectronic film deposition in the future.

In this study, HiPIMS technology is used to search for technological benefits in the field of rhenium oxide synthesis. We study the HiPIMS of a Re target by varying the pulsing frequency and the oxygen content in the discharge atmosphere. At every given sputtering condition, the current-voltage-time (*C-V-t*) characteristics, as well as the plasma optical emission spectra (OES), were recorded. At certain conditions, the rhenium oxide films were deposited and characterised by X-ray diffraction (XRD), X-ray photoelectron spectroscopy (XPS), scanning electron microscopy (SEM), UV-Vis-NIR spectroscopy, and conductivity measurements. The composition $ReO_x$ highly depends on the synthesis conditions, and $x$ ranges from 1.6 to 2.9. The deposition temperatures mainly determines the oxidation state of rhenium ions and the corresponding structure. In addition, the composition may be affected by the amount of oxygen used during sputtering. The annealing in air promotes further oxidation of rhenium oxide thin films. To the best of our knowledge, this is the first study of rhenium oxides thin films deposition by HiPIMS.

## 2. Experimental details

The experiments were performed using the R&D vacuum coater SAF25/50 (Sidrabe Vacuum, Ltd.). A two-inch circular, balanced magnetron (Gencoa, Ltd) was used. The power was supplied by a MP2-AS 200 pulse power supply unit (Magpuls GmbH) operating in the unipolar negative regime at a constant average power of 100 W. The constant power regime was selected to prevent target damage. The negative voltage was supplied to the magnetron cathode against a grounded anode. The discharge voltage and current were measured by a Magpuls GmbH measuring system and recorded by a RIGOL DS1054 digital oscilloscope.

The plasma optical emission spectra (OES) from the discharge were collected by an optical fibre probe located in the chamber overlooking the discharge parallel 1.5 cm above the target surface. The time-average OES were measured by a PLASUS EMICON MC spectrometer (200–1100 nm).

A metallic Re target (99.99% purity, 50 mm diameter, and 3 mm thickness) was sputtered in metallic (Ar) and reactive ($Ar+O_2$) atmospheres. The *C-V-t* characteristics and OES were studied as a function of the pulsing frequency (110 Hz–952 Hz and DC) by changing the time between 50 μs long pulses and the oxygen gas flow rate (0–10 sccm). In all cases, the pulse voltage was constant for the entire pulse length. The arc-detect threshold value was set to 50 A in the HiPIMS mode. Before the investigation of the sputtering process and the $ReO_x$ film deposition, the chamber ($\approx 0.1$ $m^3$) was pumped down to a base pressure below $9.5 \times 10^{-6}$ mbar by a turbo-molecular pump backed with a rotary pump. The Ar (99.99% purity) flow rate of 20 sccm was kept constant in both the metallic and reactive modes. The pumping speed was altered by a throttle valve to set the working pressure of approximately $7.3 \times 10^{-3}$ mbar during the sputtering.

The film thickness was determined by a CART Veeco Dektak 150 profilometer. The structure of the samples was examined by a Rigaku MiniFlex 600 X-ray diffractometer with Cu Kα radiation. The morphology of the $ReO_x$ films was characterized by a Thermo Scientific Helios 5 UX high-resolution dual-beam microscope.

X-ray photoelectron spectroscopy (XPS) analyses were carried out using a ThermoFisher ESCALAB Xi+ instrument using a monochromatic Al Kα X-ray source. The instrument binding energy scale was calibrated to give binding energy at 932.6 eV for the Cu $2p_{3/2}$ line of a freshly etched metallic copper. The charge compensation system was used. The surface of each sample was irradiated with a flood of electrons to produce a nearly neutral surface charge. The spectra were recorded using an X-ray beam size of 900 × 10 μm with a pass energy of 20 eV, and a step size of 0.1 eV. Data from all materials were referenced using the main signal of the carbon 1s spectrum occurring at 284.8 eV. The carbon 1s spectrum was collected using high-energy-resolution settings. No cleaning of the surface (etching) was used before measuring the survey and high-resolution spectra, as higher-order rhenium–oxygen compounds are easily reduced during sputtering. Mixed Gaussian-Lorentzian peak shapes were used to fit the Re 4f XPS high-resolution spectra by the XPSPEAK41 software. The spacing between the $4f_{5/2}$ and $4f_{7/2}$ components for all peaks was fixed to correspond to the spin-orbit splitting of 2.5 (+0.2/-0.1) eV. To improve the reliability of the peak fitting, the intensity ratio of $4f_{5/2}$:$4f_{7/2}$ peaks was set to 0.787.

The electrical resistivity was measured in the Van der Pauw configuration using a Hall effect system, HMS5000. The film's transmittance and reflectance in the range of 330–2500 nm were determined by an Agilent Cary 7000 spectrophotometer. The sample was placed at an angle of 6° against the incident beam, and the detector was placed at 180° behind the sample to measure the transmittance and at 12° in front of the sample to measure the specular reflectance.

## 3. Results and discussion

### *3.1. Discharge waveforms*

The HiPIMS process of a Re target was investigated in both metallic and reactive modes before the rhenium oxide film deposition. In the beginning, the Re target sputtering in an Ar atmosphere was gradually changed from a DC to a low-frequency pulsed regime (110 Hz) by increasing the off-time ($t_{off}$) between pulses (up to 9 ms) to study the influence of the pulsing frequency on the magnetron discharge. The average power of 100 W and the pulse duration time of 50 μs were kept constant.

The peak current and the voltage grow when the frequency is reduced to maintain the average power (Fig. 1(a)). The sputtering regime reaches the HiPIMS level of 0.5 $kW/cm^2$ (defined in [32]) in the $t_{off}$ range between 3 and 6 ms. There is a time lag with respect to the applied voltage of approximately 10 μs before the current starts to rise. The delay time is associated with the discharge ignition phase where the current is very small and the plasma is negligible. It seems that the delay time with $t_{off}$ does not change much. This is understandable because a time lag should decrease with voltage; on the other hand, it should increase if fewer charged particles are available at the beginning of a pulse, which is the case when the pulse repetition rate is reduced [33]. After the time lag, the current increases sharply because the production of electrons is faster compared to the recombination and disappearance [34]. The increase in the energetic electrons, which are trapped in the magnetic field close to the target surface, strongly enhances the working gas ionisation. The plasma impedance decreases as the concentration of the charge carriers grow. The initial phase of the current rise is dominated by gas ions [35], and electrons gain energy from sheath acceleration and Ohmic heating [36, 37]. Also, working gas rarefaction sets in at this point in time. The rarefaction is a localised depletion of the working gas in front of a target due to gas heating, which is caused by the discharge itself, the hot target surface, the momentum transfer between sputtered atoms (also reflected sputtering gas atoms) and gas, and the electron impact ionisation [38, 39].

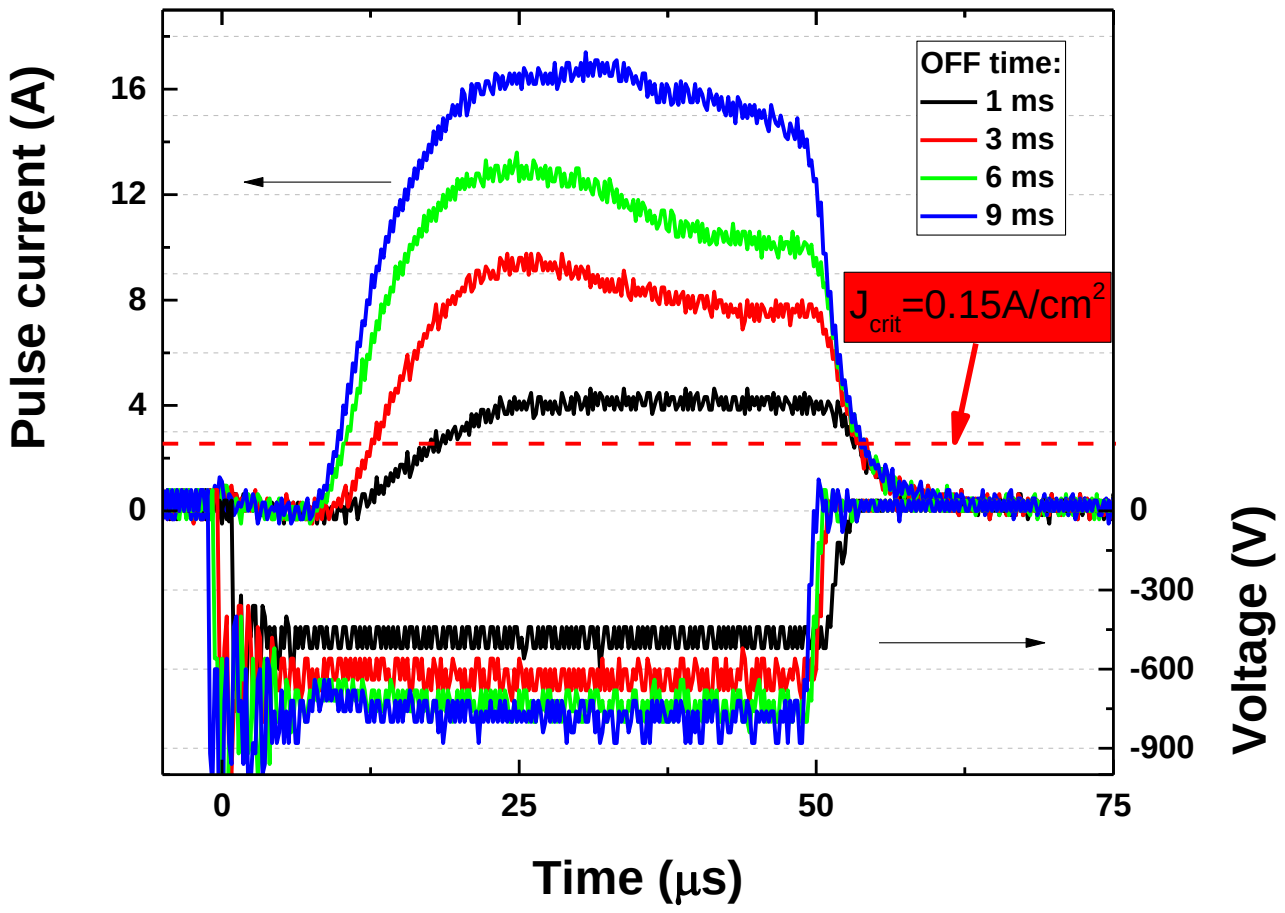

Fig. 1. *C*-*V*-*t* characteristics as a function of frequency during the Re target sputtering in an Ar atmosphere at a constant average power mode (100 W) and pulse length (50 μs).

The current reaches the peak value of 4 A (the peak current density of 0.24 A/cm$^2$) in the middle of the pulse (at 25 μs) and remains constant for the remaining pulse time when using a $t_{off}$ of 1 ms. When $t_{off}$ is increased above 1 ms, the shape of the current profile changes qualitatively. After the peak current in the middle of a pulse, the current starts to decrease. The depletion of working gas atoms in the ionisation region gets stronger and faster when the discharge current rises. Consequently, the source of ions that carry the discharge current is highly reduced when the current reaches the peak, resulting in a rapid decrease in the current after the peak [40]. The temporal evolution of the discharge current reaches a stable plateau value of 8 A (0.48 A/cm$^2$) at the end of a pulse when using a $t_{off}$ of 3 ms. It is reasonable to assume that a stable plateau current would also be observed at a longer $t_{off}$ (6 and 9 ms in Fig. 1) if longer pulses are used. The highest peak current of 17 A (1.02 A/cm$^2$) is measured when using a $t_{off}$ of 9 ms. In this case, the peak value is reached later into the pulse at 31 μs compared to the shorter $t_{off}$ values.

The critical current density $J_{crit}$ (defined in [41]) is approximately 0.15 A/cm$^2$ in our case. This value defines the maximum possible ion current if all incoming gas atoms from the surrounding volume are singly ionised and accelerated towards the target, taking into account the thermal refill rate of the working gas. A higher current density than $J_{crit}$ cannot be obtained without either an additional working gas or sputtered target material ion recycling ($I_{gas\text{-}recycle}$ and $I_{SS}$, respectively) or a mixture of both processes. This means that already at a $t_{off}$ of 1 ms, a significant fraction of the discharge current ($I_D$) carrying ions is recycled. In general, the fractions $I_{gas\text{-}recycle}/I_D$ and $I_{SS}/I_D$ are determined by the self-sputter yield $Y_{SS}$ [42]. For high-$Y_{SS}$ target materials, $I_{SS}/I_D$ dominates over $I_{gas\text{-}recycle}/I_D$. For example, a HiPIMS discharge of an Al target with $Y_{SS}$ = 1.1 at 600 V is strongly dominated by the self-sputtering recycling where a large amount of sputtered atoms is ionised (ionisation energy of neutral Al is around 6 eV) and drawn back to the target, creating a high discharge current [36]. Since $Y_{SS}$ of Re is close to that of Al, approximately 1.1 at an even lower voltage of 480 V (value of the voltage at 1 ms $t_{off}$ in this study) [43], it can be argued that the self-sputtering dominates over working gas ion recycling also in the case of Re. Although the ionisation energy of neutral Re (7.8 eV) is higher than that of Al, it is still well below the ionisation energy of neutral Ar (15.8 eV). Since the Ar ion contribution is small and singly ionised Re cannot generate secondary electrons, Ohmic heating is the main source for electron energisation [44]. The above-described analysis of the temporal evolution of a discharge current during the Re target sputtering confirms the gas-sustained self-sputtering regime where the current is dominated by Re ion recycling. At the same time, an ionized Ar gas is required and provides an important contribution to the discharge. In addition, the discharge current at the end of the pulse gradually increases with the voltage without a sudden large jump (Fig. 1). This also indicates the gas-sustained self-sputtering since the self-sputter yield is a smooth function of the energy of the bombarding ions.

The evolution of the discharge current during reactive (r-) HiPIMS is shown in Fig. 2. The shape of a current waveform changes and the peak value reduces significantly immediately after the addition of oxygen gas to the background gas. In general, the peak current tends to increase when a compound layer starts to form on the target surface in r-HiPIMS [29, 45]. This is due to a reduced sputter yield with a consequential increase in the working-gas recycling to compensate for the reduced self-sputtering. As a result, the secondary electron contribution fraction into the electron energisation also increases, and the current can reach very high values [46]. However, this is not the case in our study. The current waveforms in r-HiPIMS of Re indicate that the discharge is still in the self-sputter recycling mode, but with a lower peak current compared to the pure metal mode. This indicates a lower ionization efficiency of Ar gas and sputtered Re atoms. This might be explained by lower secondary electron emission. The work function of 5.1 eV for a pure Re surface substantially increases to approximately 7.1 eV after oxidation [47]. A higher work function leads to lower secondary emission according to the empirical formula for the emission determined by the potential energy of the arriving ion projectiles [48]. The increase from 10 A to 12 A observed in the peak current with the oxygen flow (Fig. 2) indicates an increase in the gas-recycling contribution.

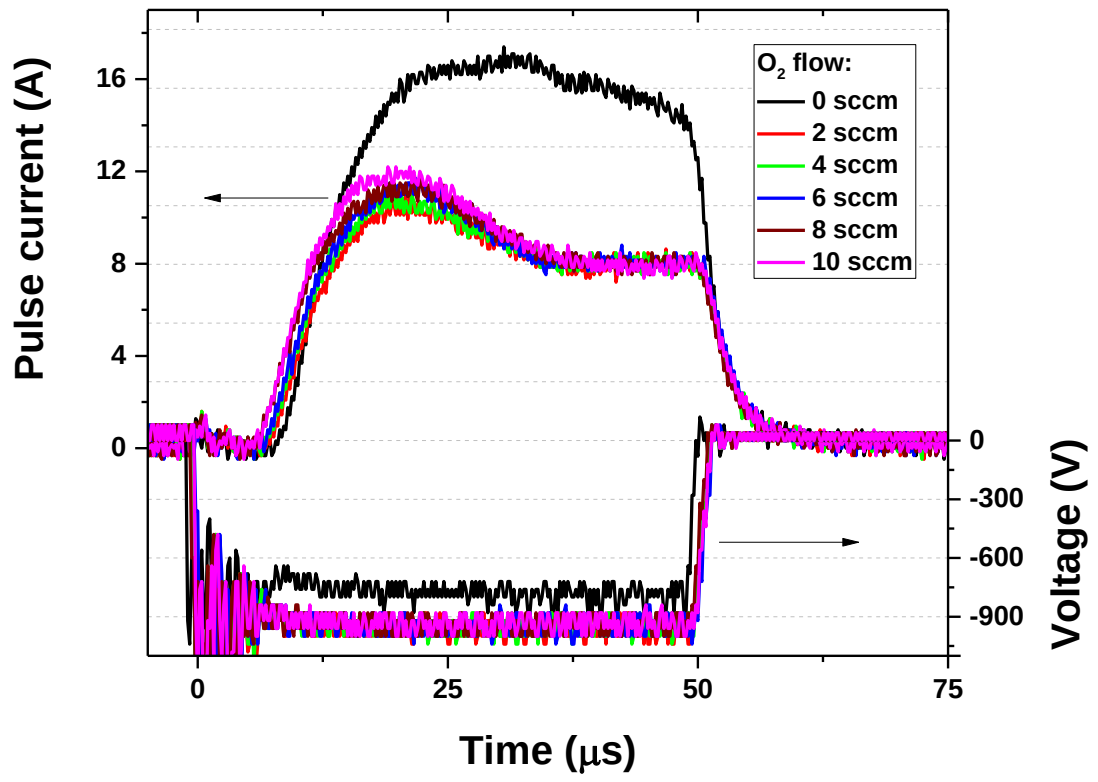

**Fig. 2.** *C-V-t* characteristics as a function of oxygen flow during the Re target sputtering in an Ar and $O_2$ atmosphere at a constant average power mode (100 W) and pulse length (50 μs).

## *3.2. Plasma OES*

The plasma composition was investigated using OES recorded approximately 1.5 cm above the Re target surface. The spectra as a function of $t_{off}$ and oxygen gas flow are shown in Fig. 3. The plasma emission is dominated by neutral Ar (Ar I) emission lines with relatively less intense lines of neutral Re (Re I) at the DC discharge. Low intense ionized Ar (Ar II) lines can also be observed. The intensity of the Ar I and Ar II lines rapidly decreases, whereas the intensity of Re I increases sharply when the discharge is switched to the pulsed regime with a $t_{off}$ of 1 ms. This is the result of the already described Ar gas rarefaction. A further increase in $t_{off}$ reduces the intensity of the Ar I, Ar II, and Re I lines (Fig 4(a-c)), whereas the intensity of the Re II lines grows (Fig 4(d)), which is in line with the evolvement of self-sputter recycling. However, the intensity of the Re II lines is still significantly lower compared to the Re I lines. At the low frequency discharge (110 Hz, $t_{off}$ of 9 ms), the plasma emission spectrum mainly consists of the Re I emission lines.

The ionised flux fraction (the ratio of the ion flux to the total particle flux arriving at the substrate or detector [20]) in the case of Al and Ti is ≥ 50% for typical HiPIMS conditions – a peak current density of approximately 1 A/cm$^2$ and a sputtering pressure in the range from 0.5 to 2.0 Pa [45]. It can be predicted that the ionisation fraction in the case of Re could have a similar value because, although Re has a higher ionisation energy (7.8 eV) compared to Al (6.0 eV) and Ti (6.8 eV), it has a larger collision cross-section of electron impact ionisation [49].

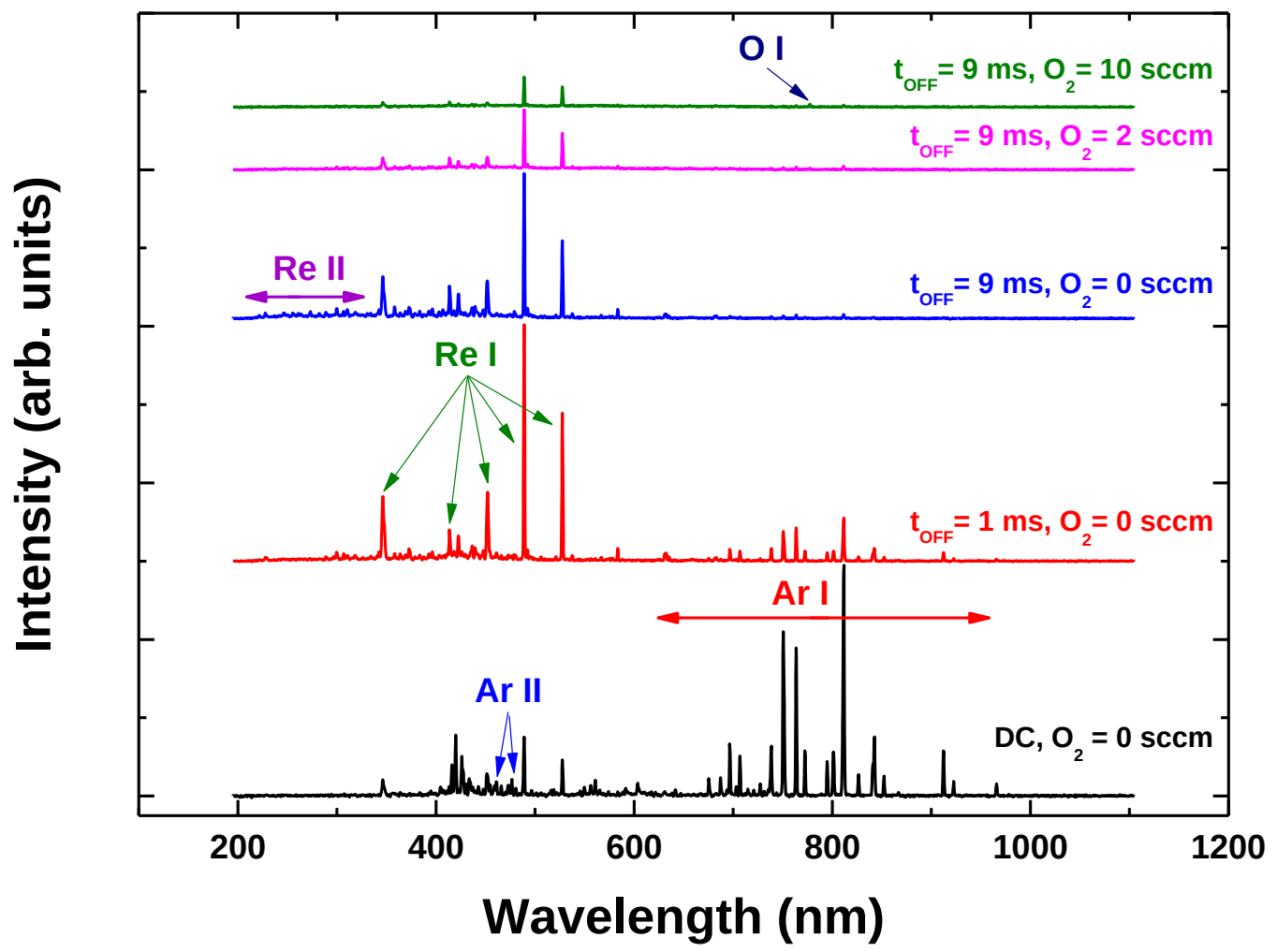

**Fig. 3.** OES spectra as a function of $t_{\mathrm{off}}$ and oxygen flow rate during the Re target sputtering at a constant average power mode (100 W) and pulse length (50 μs).

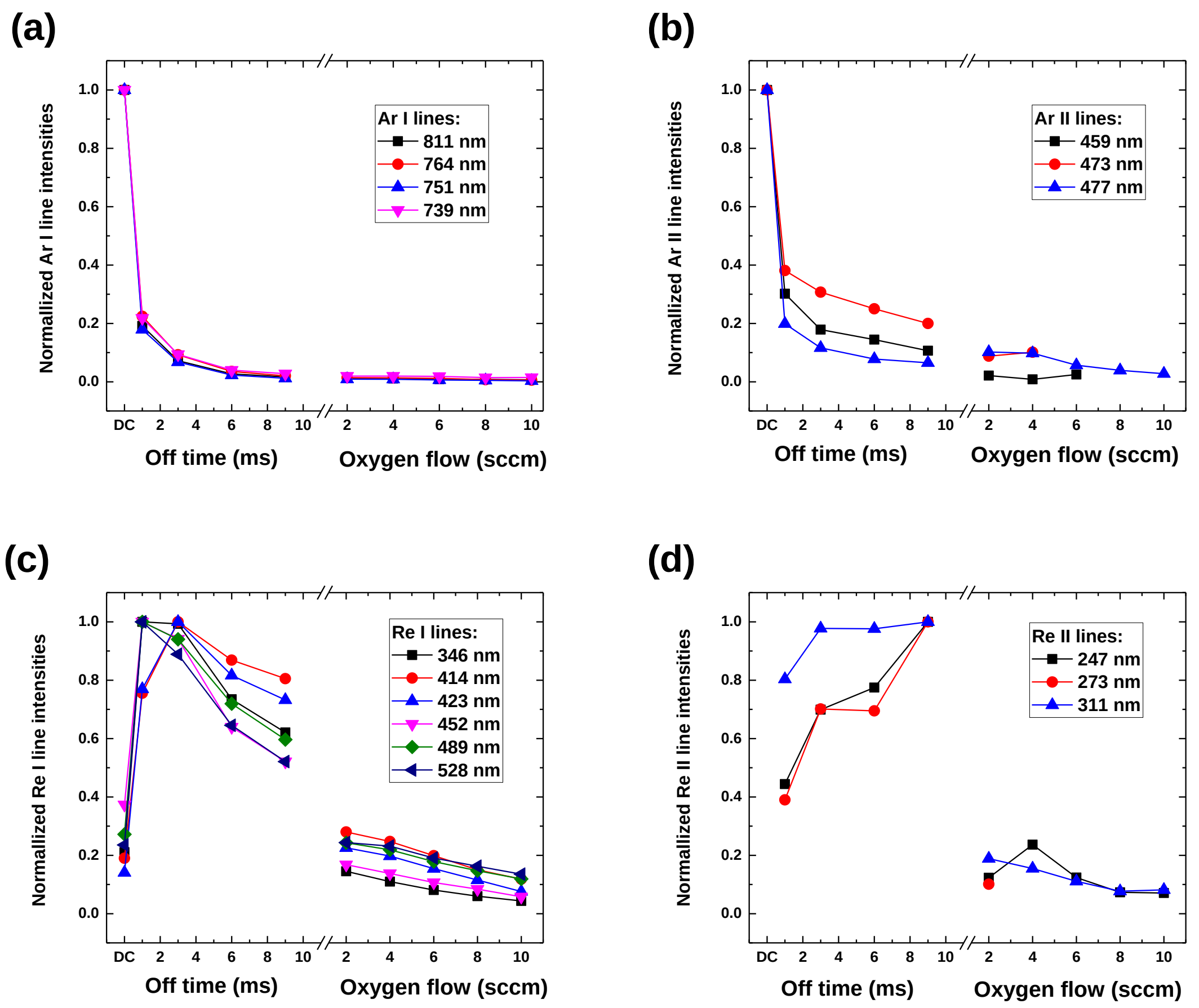

**Fig. 4.** Normalised emission line intensities of neutral Ar (a), single ionised Ar (b), neutral Re (c), and single ionised Re (d) as a function of $t_{off}$ and oxygen flow. The highest detected intensity of each line was normalised to 1.

### *3.3. Prepared $ReO_x$ samples*

$ReO_x$ ($x \approx 1.6$–$2.9$) thin films were deposited on fused quartz substrates by reactive high power impulse magnetron sputtering (r-HiPIMS) in a mixed $Ar/O_2$ atmosphere. The substrates were ultrasonically cleaned with acetone and 2-isopropanol for 15 min each, rinsed with distilled water, and then dried under blown $N_2$ gas. Samples were produced at three different substrate temperatures: 36 °C (without intentional heating), 150 °C, and 250 °C; and at two different oxygen-to-argon flow ratios: 0.4 and 0.5 (see Table 1). The substrates were located approximately 13 cm away from the target surface (a substrate facing the target axis) on the grounded substrate holder. The angle between the target and the substrate was fixed at ≈ 45°. After the depositions, the samples were directly transferred (without breaking the vacuum) and stored in an Ar atmosphere glove-box connected to the vacuum system to protect the films from possible degradation (see Fig. S1 in the supplementary material) under ambient conditions. The average thickness of the films is approximately 150 nm. During the annealing of the films at 250 °C in the air for 3 h, a clean piece of quartz was placed on the sample surface to avoid the evaporation of rhenium oxide. The resistivity of the films is on the order of $10^{-3}$ Ωcm and summarised in Table 1. No clear correlation was observed between the resistivity and the deposition parameters or post-treatment. The measurements of the as-deposited and annealed films were performed for two separate pieces of the same sample. There may be slight differences between the pieces, which may cause inconsistency.

Table 1

Oxygen-to-argon flow ratio and substrate temperature during the deposition of $ReO_x$ samples by r-HiPIMS. The table also contains data on sample thickness (nm) and resistivity (Ωcm).

| Sample | $O_2$/Ar flow ratio | Substrate temp. (°C) | Thickness (nm) | Atomic composition Re/O | | Resistivity ($10^{-3}$ Ωcm) | |
|---|---|---|---|---|---|---|---|
| | | | | As-deposited | Annealed | As-deposited | Annealed |
| *a* | 0.40 | 37 (RT) | 142 | 0.48 | 0.46 | 0.9 ± 0.3 | 1.1 |
| *b* | 0.50 | 36 (RT) | 183 | 0.45 | 0.35 | 1.5 | 1.2 |
| *c* | 0.40 | 150 | 133 | 0.58 | 0.55 | 2.2 | 1.1 |
| *d* | 0.50 | 150 | 152 | 0.58 | 0.57 | 0.8 | 1.1 |
| *e* | 0.50 | 250 | 148 | 0.61 | 0.54 | 1.9 | 4.2 |

### *3.4. Structural analysis using XRD*

X-ray diffraction patterns of the as-deposited $ReO_x$ films on fused quartz substrates (recorded over a 2θ-range of 10°–80°) are shown in Fig. 5 depending on the deposition temperature and the oxygen-to-argon flow ratio. The broad peak at 2θ=21.2° is due to the substrate. X-ray patterns of the films grown on non-heated substrates (samples *a* and *b*) do not show Bragg peaks associated with any of the pure rhenium oxides suggesting that the films are X-ray amorphous. A peak at around 16.6° with low intensity corresponds to crystalline perrhenic acid monohydrate ($HReO_4 \cdot H_2O$) [16]. Interestingly, this phase was not observed in rhenium oxide films, which were deposited at RT by reactive pulsed-DC magnetron sputtering in our earlier study [14]. However, it formed after the same annealing procedure was used in this study. Two XRD peaks with relatively low intensity at 37.3 and 53.9° appear when the deposition temperature increases up to 150 °C or 250 °C (samples *c*, *d*, and *e*). These peaks correspond to the orthorhombic $\beta$-$ReO_2$ lattice planes (200) and (221) according to the powder diffraction card ICDD 00-009-0274. Note that the signal from $HReO_4 \cdot H_2O$ disappears at elevated substrate temperatures. The size of the crystallites according to the Scherrer equation is about 7 nm. The only difference observed between 150 °C or 250 °C deposition temperatures for the same oxygen-to-argon ratio is the intensity ratio between the two XRD peaks. The oxygen-to-argon ratio, which ranged from 0.4 to 0.5 in this study, did not affect the structure of the as-deposited films according to the XRD measurements. In the framework of this study, it was not possible to obtain a crystalline $ReO_3$ film during the deposition process.

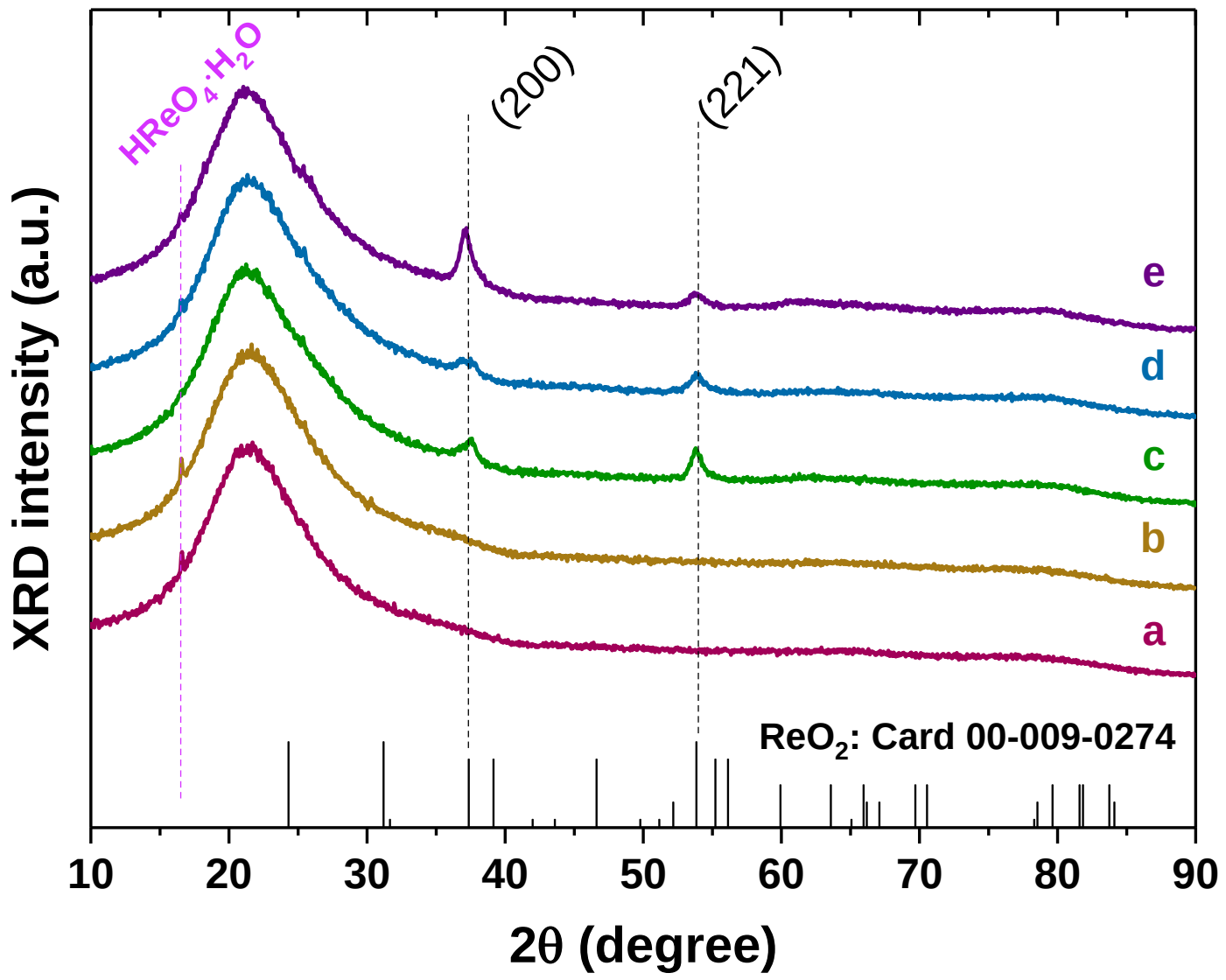

**Fig. 5.** X-ray diffraction patterns of the as-deposited $ReO_x$ films produced by r-HiPIMS at different substrate temperatures and oxygen-to-argon flow ratios. The deposition parameters of samples *a*-*e* can be found in Table 1. The angles of the Bragg peaks for the orthorhombic β-$ReO_2$ [50] and tetragonal $HReO_4$ [16] phases are shown by vertical black and purple dashed lines, respectively.

The as-deposited films were annealed as described in the experimental details. The X-ray diffraction patterns in Fig. 6 reveal the formation of a crystalline $ReO_3$ phase in the X-ray amorphous films (samples *a* and *b*) after the annealing. Diffraction maximums at 23.8°, 33.9°, 41.8°, 48.6°, 54.8°, 60.5°, and 76.4° correspond to the (100), (110), (111), (200), (210), (211), and (300) planes of the cubic $ReO_3$. Rietveld analysis of the XRD data gives the lattice parameter of 3.75 Å, which is in excellent agreement with the ICDD card 04-004-8088. The conversion process from the amorphous to the crystalline phase takes a longer time (at least 2 h) compared to the pulsed-DC deposited films (30 min) in Ref. [14]. This might be since films grown by HiPIMS compared to DC magnetron sputtering are, in general, denser, which hinders the crystallisation process. In addition, the crystallite size of 50 nm according to the Scherrer equation is also smaller compared to the pulsed-DC deposited films (100–500 nm). After the annealing, the crystalline $ReO_3$ phase was also observed in the film deposited at 150 °C with an oxygen-to-argon ratio of 0.4 but not in the film with an oxygen-to-argon ratio of 0.5. Since the resistivity (Table 1) of the films containing the crystalline $ReO_3$ phase is about two orders of magnitude higher than theoretically and experimentally obtainable ($0.9 \times 10^{-6}$ Ωcm [51, 52]), it seems that the whole structure is not converted to crystalline $ReO_3$ and contains structural defects, thus the mobility and concentration of the charge carriers are reduced. No changes in the X-ray diffraction pattern of the film deposited at 250 °C were observed after the annealing procedure.

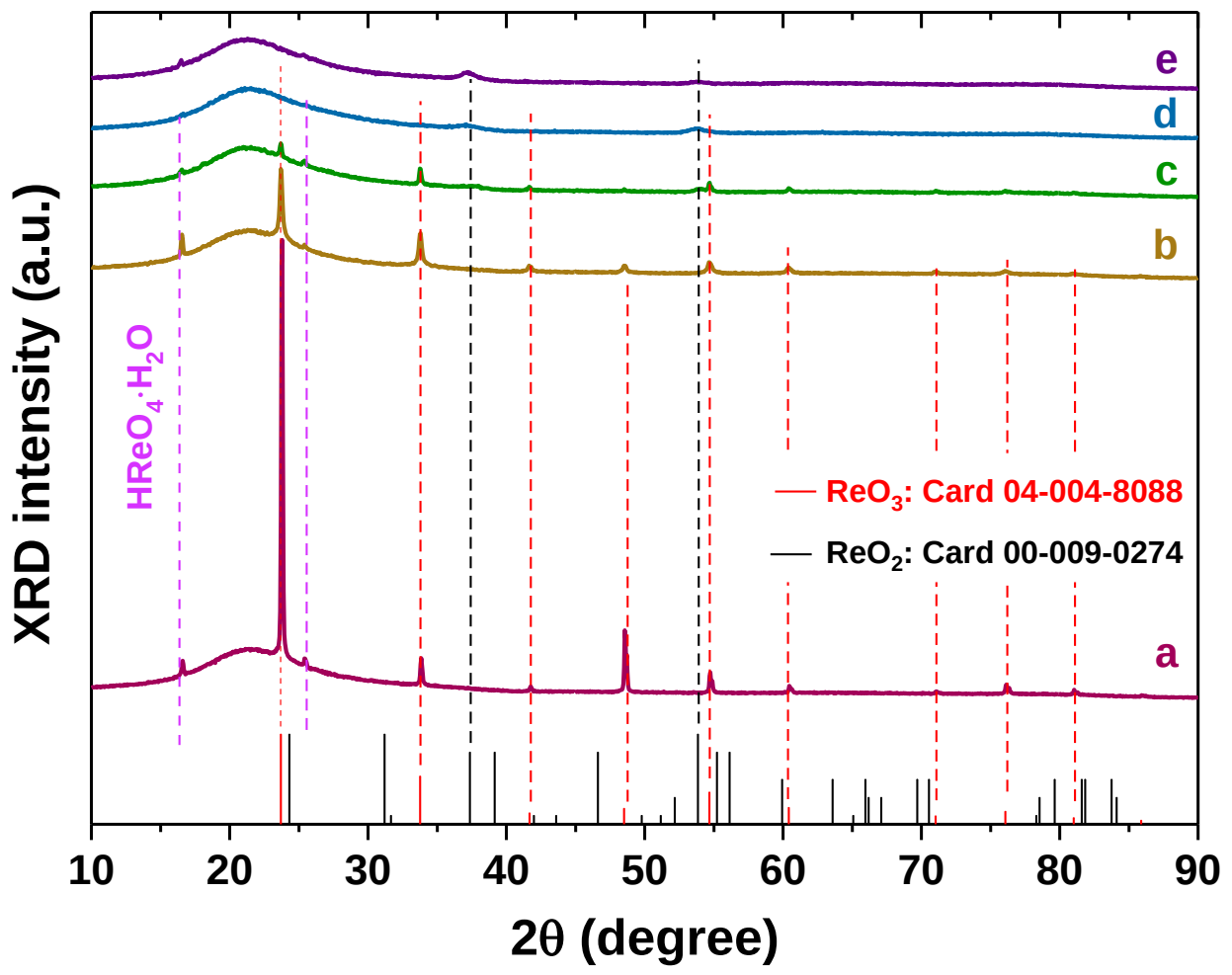

**Fig. 6.** X-ray diffraction patterns of the annealed (250 °C in the air for 3 h) $ReO_x$ films deposited by r-HiPIMS at different substrate temperatures and oxygen-to-argon flow ratios. The deposition

parameters of samples *a*-*e* can be found in Table 1. The angles of the Bragg peaks for the cubic $ReO_3$ [8], orthorhombic β-$ReO_2$ [50], and tetragonal $HReO_4$ [16] phases are shown by vertical red, black, and purple dash lines, respectively.

*3.5. Surface morphology – SEM images*

The surface of the films before and after the annealing was imaged by SEM (Fig. 7 and Fig. 8). The as-deposited films at RT (samples *a* and *b*) have a defect-rich surface. The film deposited at an oxygen-to-argon ratio of 0.4 contains large open cavities with a size of a few hundred nanometres (Fig. 7(a)), which are most likely formed after exposure to air. When the ratio is increased to 0.5, the morphology of the surface differs, but it is still porous with a large number of submicron voids (Fig. 7(b)). The surface gets significantly smoother and is featureless when the substrate temperature is increased (Fig. 7(c–e)). The films deposited at 150 °C (samples *c* and *d*) have a dense surface with nano-sized open cavities. The surface becomes even denser with a fine-grained structure for the film deposited at 250 °C (sample *e*).

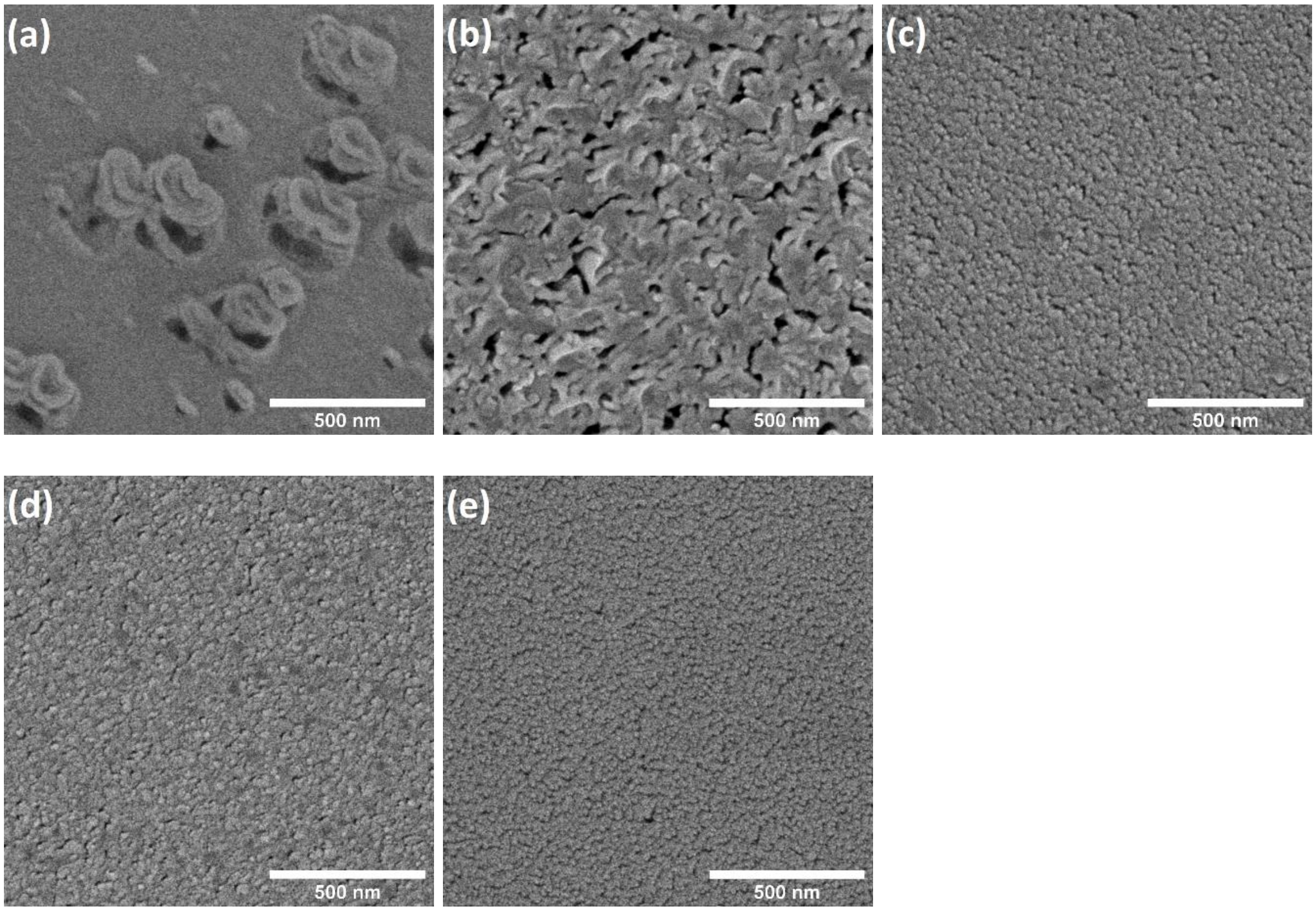

**Fig. 7.** SEM images of the as-deposited rhenium oxide films deposited by r-HiPIMS at RT and $O_2$/Ar = 0.4 (a), RT and $O_2$/Ar = 0.5 (b), 150 °C and $O_2$/Ar = 0.4 (c), 150 °C and $O_2$/Ar = 0.5 (d), 250 °C and $O_2$/Ar = 0.5 (e).

After the annealing, the films deposited at RT contain crystallites of different sizes (on average ≈ 50 nm) and orientations (Fig. 8(a–b)). The crystallisation is in line with our XRD measurements that show a crystalline $ReO_3$ phase. The surface structure is not closely packed, most likely due to the partial sublimation of $ReO_3$ during the annealing [53].

Crystallites of sub-micrometre size are observed on the surface of the films deposited at 150 °C after the annealing (Fig. 8(c–d)). The crystallites are most likely of the insulating solid perrhenic acid monohydrate $HReO_4 \cdot H_2O$ phase and form on the surface over time after exposure to air moisture [3, 14]. It has been shown that the rhenium-to-oxygen concentration ratio on the surface of rhenium oxide deposited films decreases over time [54]. The X-ray diffraction patterns do not show any narrow peaks corresponding to the single crystals, most likely because the SEM measurements were performed later than the XRD. After half a year, the X-ray diffraction pattern of the annealed sample *c* shows only the $HReO_4 \cdot H_2O$ phase, and all the peaks due to $ReO_3$ have disappeared (Fig. S2 in the supplementary material). It also clearly shows the instability of the $ReO_3$ phase under ambient conditions. The surface of the crystalline $ReO_3$ phase revealed by XRD (Fig. 6) is, again, not densely packed for the film deposited at 150 °C with an oxygen-to-argon ratio of 0.4 (Fig. 8(c)). For an $Ar/O_2$ ratio of 0.5, the surface is densely packed, and the partly immersed crystallites on the surface have a tetragonal dipyramidal shape with sharp edges, which corresponds to the $I4_1/a$ space group of $HReO_4 \cdot H_2O$. The surface of the sample deposited at 250 °C still has a dense fine-grained structure without significant changes after the annealing.

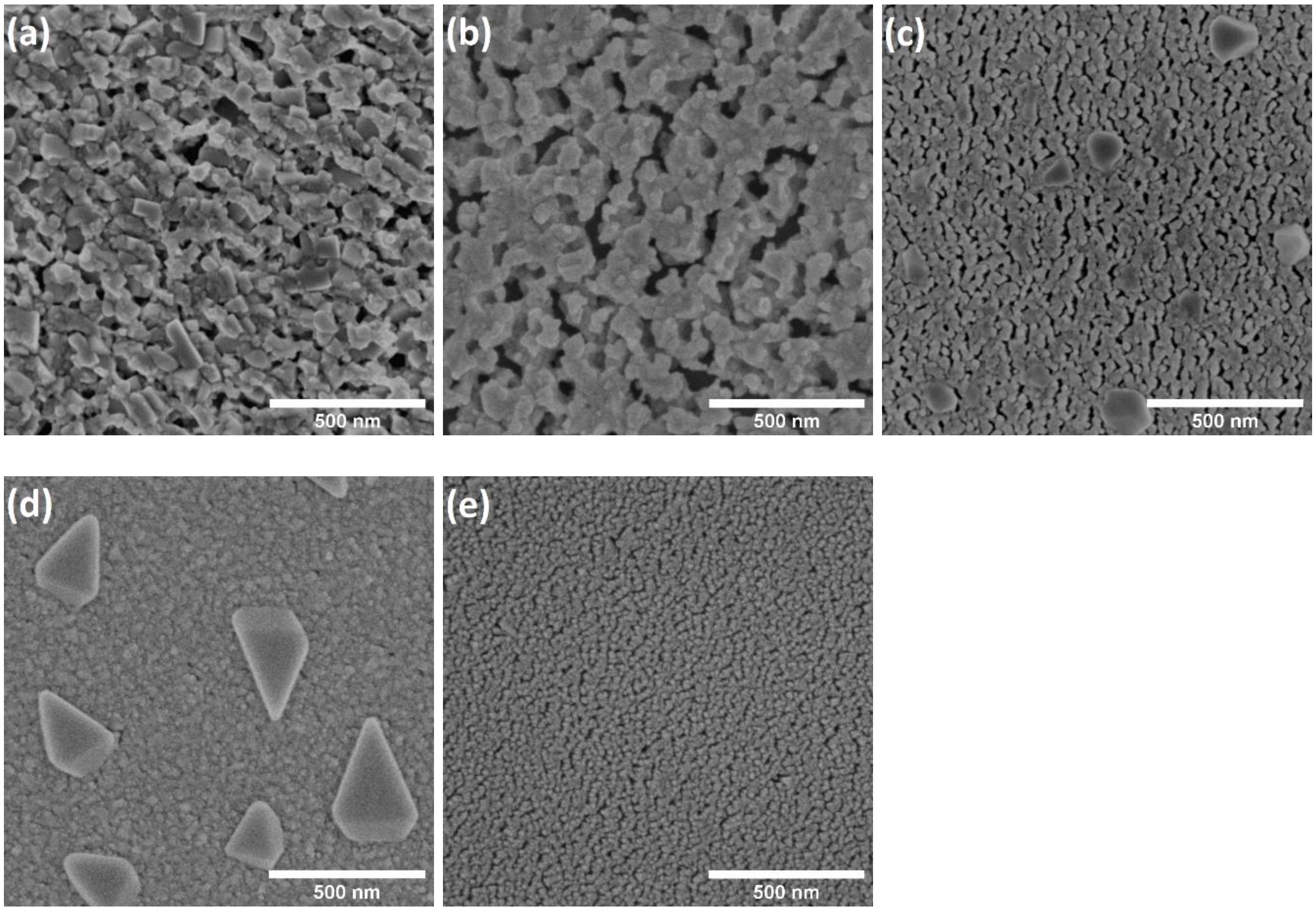

**Fig. 8.** SEM images of the annealed (250 °C in air for 3 h) rhenium oxide films deposited by r-HiPIMS at RT and $O_2$/Ar = 0.4 (a), RT and $O_2$/Ar = 0.5 (b), 150 °C and $O_2$/Ar = 0.4 (c), 150 °C and $O_2$/Ar = 0.5 (d), 250 °C and $O_2$/Ar = 0.5 (e).

### *3.6. Surface composition analysis – XPS measurements*

The composition and chemical state of the rhenium oxide film surfaces were studied by X-ray photoelectron spectroscopy (XPS). The measurements were performed after brief exposure of the samples to air. The survey spectra contain well-defined signals from Re and O elements (Fig. S3 in the supplementary material). The ratio of the atomic concentration between Re and O as a function of the deposition temperature and the oxygen-to-argon flow ratio in both as-deposited and annealed films is shown in Fig. 9. The Re/O ratio of the films deposited at RT (samples *a* and *b*) is about 0.46 before the annealing. Interestingly, the composition of the film deposited at RT and $O_2$/Ar = 0.5 changes significantly after the annealing and is close to $ReO_3$. This ratio grows and is in the range between 0.58 and 0.61 when the elevated deposition temperatures (150 or 250 °C) are used during the deposition (samples *c*, *d*, and *e*). The sub-stoichiometric oxide composition may explain the growth of $Re_2O_7$ and the consequential $HReO_4{\cdot}H_2O$ crystallites on the surface (Fig. 8(c,d)), as it has been shown that $Re_2O_7$ forms on the surface of the Re metal in an oxidizing environment [3]. The Re/O ratio decreases for all annealed films due to the elevated oxidation in air. In addition, the oxygen content in the films is slightly larger if a higher oxygen-to-argon flow ratio is used, which was already expected.

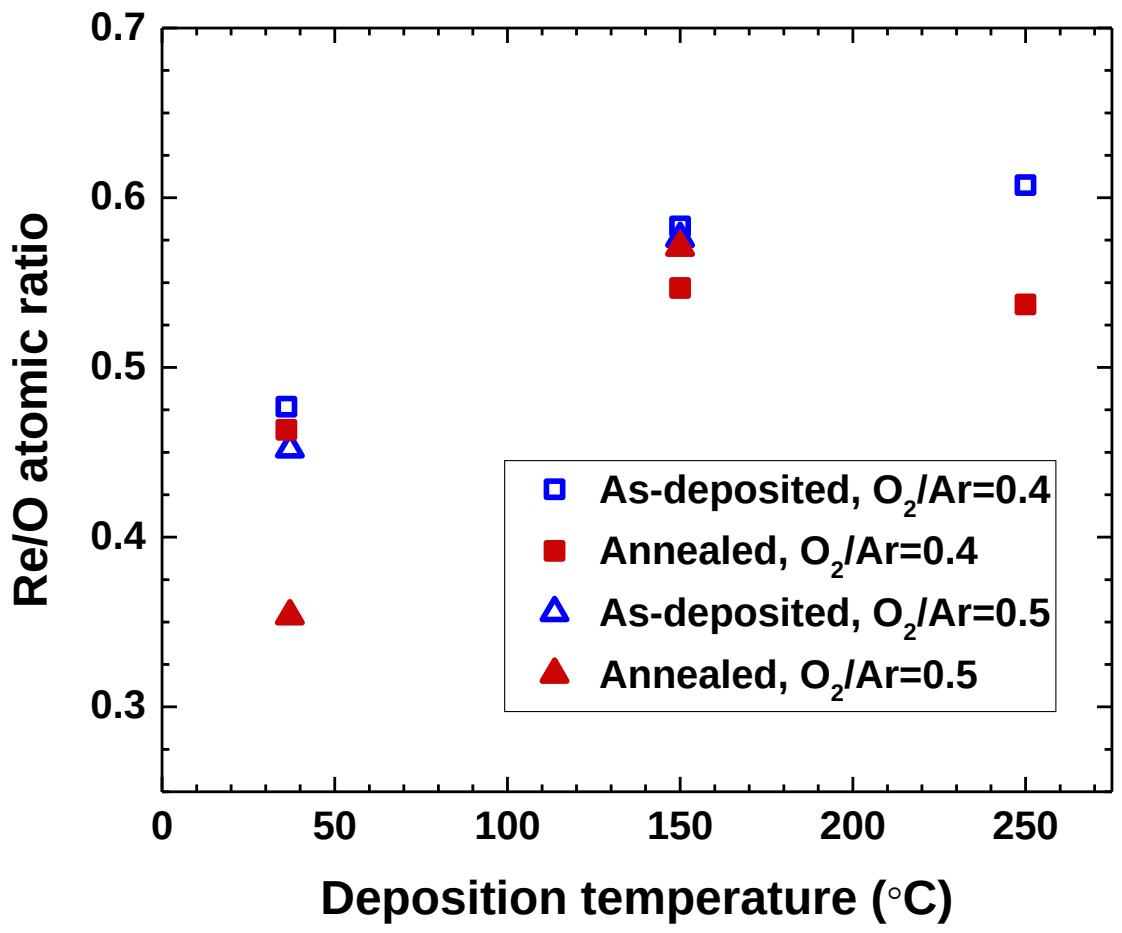

**Fig. 9.** Re to O atomic concentration ratio as a function of the deposition temperature and the oxygen-to-argon flow ratio in both as-deposited and annealed films.

The high-resolution XPS spectra of Re 4f transitions for the samples deposited at RT and 250 °C (samples *b* and *e*) are shown in Fig. 10 and Fig. 11. The influence of the deposition temperature on the shape of the spectra is visible. The spectra are complex and difficult to analyse because they consist of overlapping signals from several possible Re oxidation states (e.g. 4+, 6+, 7+ [1]) and plasmon satellites [55] within a relatively narrow energy range. There is still inconsistency in the literature about the exact binding energies for different Re oxidation states which are required to correctly identify them. This is most likely due to the different oxidation states of rhenium when it forms oxides. Some of the oxides are unstable and can easily transform into one another, which may lead to incorrect interpretations. The intense peak around 43.4 eV in Fig. 11(c) was assigned to the $Re^{6+}$ $4f_{7/2}$ and served as a reference point for other oxidation state energies because this sample clearly showed the $ReO_3$ phase according to the XRD results and the binding energy is rather close to the value of 43.1 eV found in the literature [3].

The average binding energies for three oxidation states of rhenium are summarised in Table 2. Note that in the case of the crystalline $ReO_3$ phase, broad plasmon energy-loss satellites provide an additional contribution to the spectra [56]. These features are shifted by 2.1 eV (plasma edge) from the $Re^{6+}$ peaks and coincide with the positions of the $Re^{7+}$ peaks.

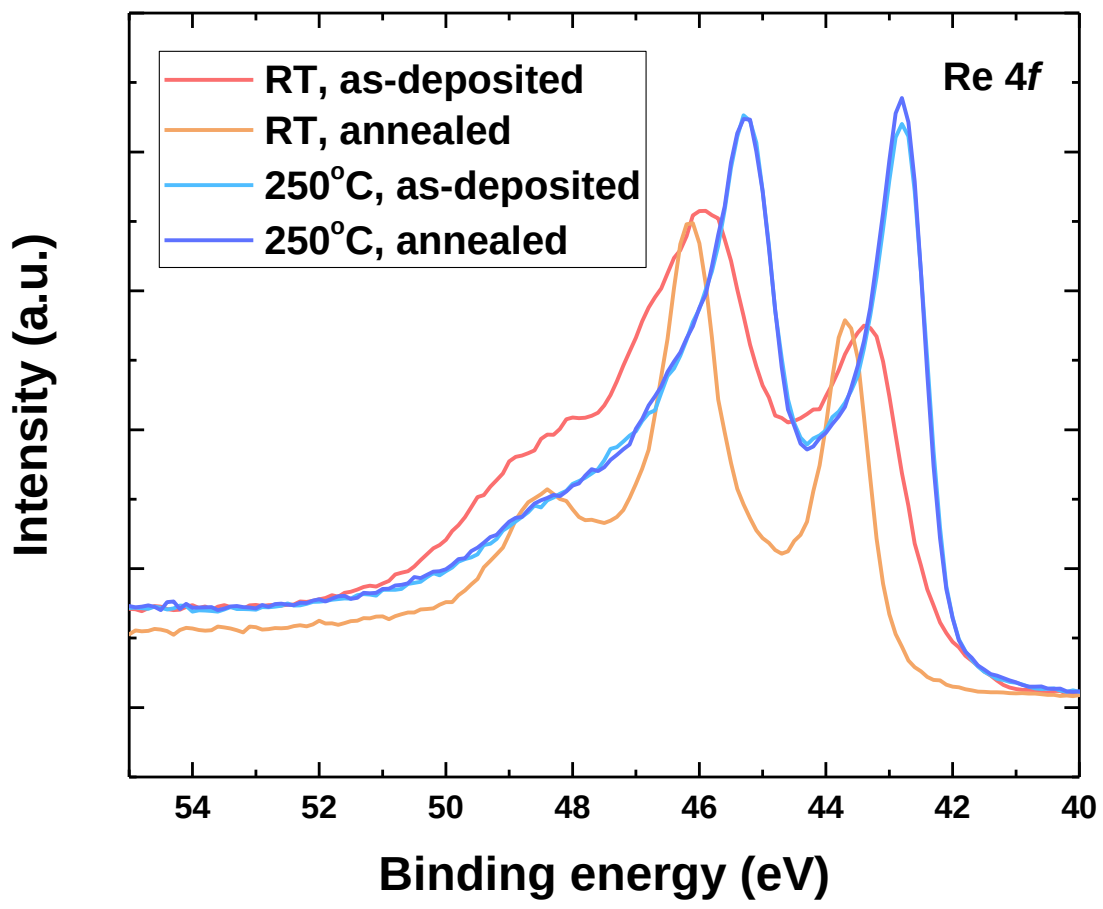

**Fig. 10.** Re 4f high-resolution XPS spectra of the rhenium oxide films deposited by r-HiPIMS at RT and 250 °C ($O_2$/Ar = 0.5 in both cases).

The Re 4f spectrum (Fig. 11(a)) of the as-deposited film at RT (sample *b*) shows that the surface mainly contains $Re^{7+}$ and $Re^{6+}$ species (68% $Re^{7+}$, 29% $Re^{6+}$, 3% $Re^{4+}$). The $Re^{4+}$ peaks disappear completely, and both $Re^{7+}$and $Re^{6+}$ peaks become narrower after the annealing, which correlates with the formation of crystalline $ReO_3$ (Fig. 11(c)). A significant contribution of the plasmon energy-loss satellites is expected here. Rhenium oxides, on the other hand, tend to form a layer of $Re_2O_7$ when exposed to air [3]. This layer then adsorbs water and forms hydroxides. As a result, the peaks towards the high-binding-energy end of the spectra (marked as 7+ here) are extremely broad and hard to resolve into individual components.

The surface of the sample deposited at 250 °C shows almost identical Re 4f spectra before and after the annealing (Fig. 11(b,d)). The spectra contain well-resolved $Re^{4+}$ peaks, which correlate with the formation of the nano-crystalline $ReO_2$ phase according to XRD. Again, after the exposure to air, the presence of $Re^{6+}$ and $Re^{7+}$ on the surface of the $ReO_2$ powder has already been demonstrated [3]. In our case, the surface contains 42% $Re^{7+}$, 33% $Re^{6+}$, and 25% $Re^{4+}$ before the annealing and 36% $Re^{7+}$, 33% $Re^{6+}$, and 31% $Re^{4+}$ after.

The high-resolution XPS spectrum of O 1s for the annealed sample *a* is shown in Fig. S4 in the supplementary material. One can see that it contains several components at 529.8 eV, 531.1 eV, and 532.8 eV. The smallest energy component corresponds to metal oxide while the other one is due to organic C-O, C=O contaminants on the surface of the sample.

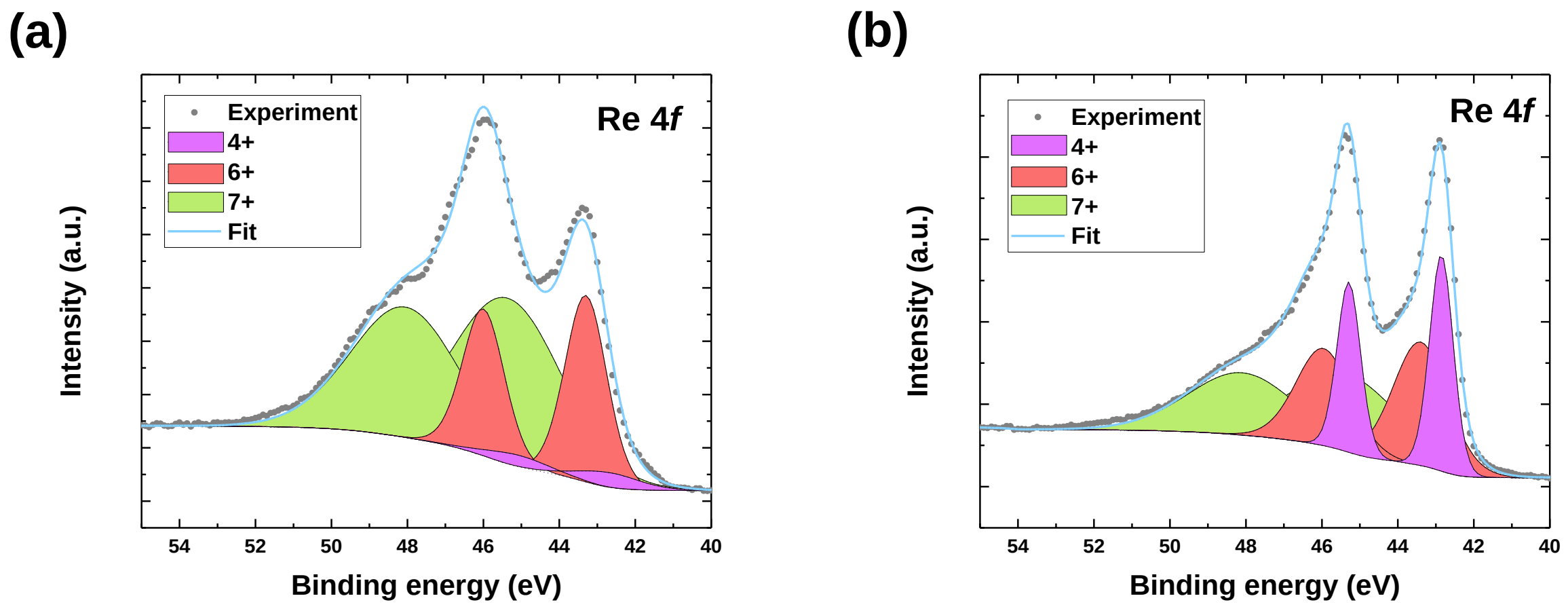

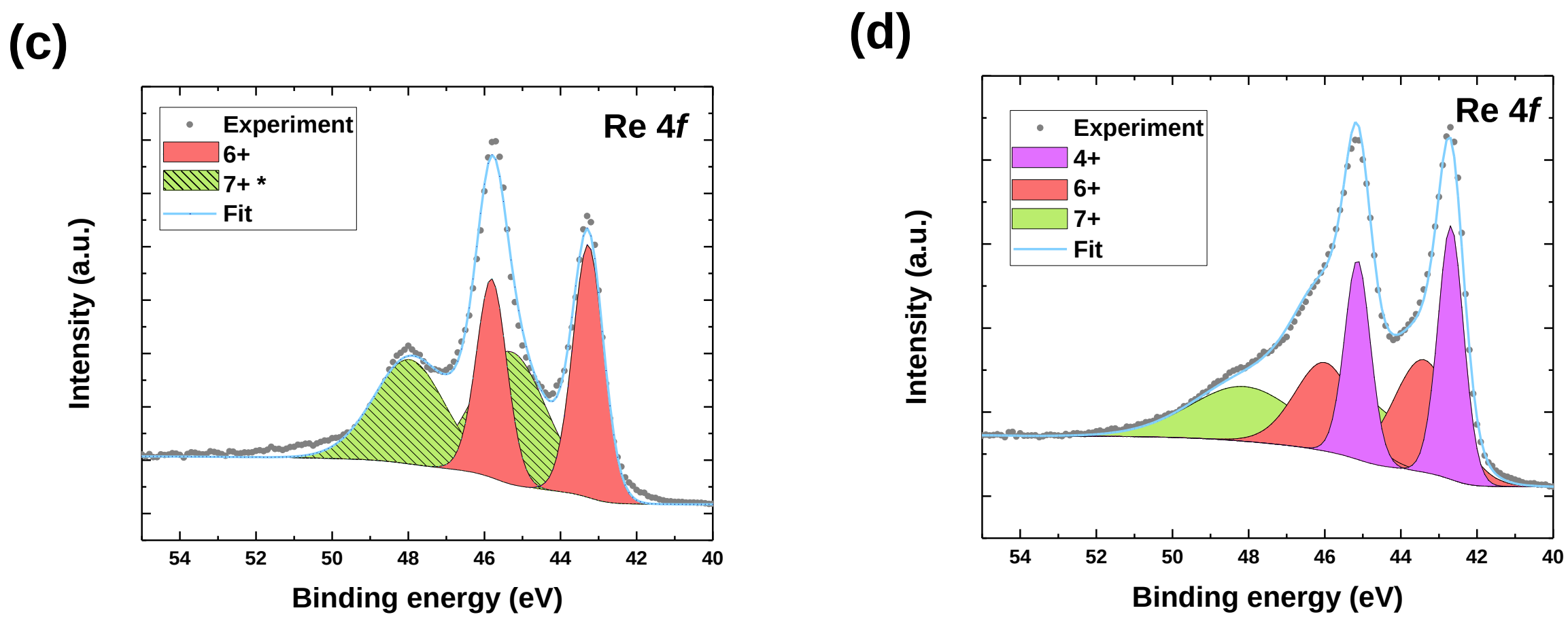

**Fig. 11.** High-resolution XPS spectra and peak fitting of Re 4f transitions in the as-deposited rhenium oxide films deposited by r-HiPIMS at RT and $O_2$/Ar = 0.5 (a), 250 °C and $O_2$/Ar = 0.5 (b), and the annealed films deposited at RT and $O_2$/Ar = 0.5 (c), 250 °C and $O_2$/Ar = 0.5 (d). The Re oxidation states are indicated in the legends: 4+ violet, 6+ red, and 7+ green (7+ * contains the contribution of plasmon energy-loss features). The grey dots represent the raw data, and the blue solid line represents the sum of all peak fits.

Table 2

Binding energies of Re $4f_{7/2}$ for different oxidation states.

| Oxidation state | Compound | Binding energy (eV) from this study | Binding energy (eV) from literature | | |
|---|---|---|---|---|---|
| | | | [3] | [54] | [57] |
| 4+ | $ReO_2$ | 42.8 | 42.2 | 42.9 | 42.5 |
| 6+ | $ReO_3$ | 43.4 | 43.1 | 44.9 | 44.9 |
| 7+ | $Re_2O_7$ | 45.4 | 45.5 | 46.9 | 46.7 |

### *3.7. UV-Vis-NIR transmittance and reflectance*

Figure 12 shows the transmittance and specular reflectance of the rhenium oxide films deposited on fused quartz measured in the range from 330 to 2500 nm. All the films absorb highly (65–95%) in the visible light range (Fig. S5 in the supplementary material). The as-deposited films exhibit a metallic dark grey colour. After annealing of the films deposited at RT (samples *a* and *b*) and the film deposited at 150 °C, $O_2$/Ar = 0.4 (sample *c*), the colour changes from grey to red in the reflected light, indicating the formation of a crystalline $ReO_3$ phase. The spectra of these samples show a transmittance window from 400 to 650 nm, where the transmittance increases to approximately 27.5% for samples *b* and *c* (Fig. 12(c)). Within this range, the reflectance reaches a minimum value of approximately 1% at the plasma edge of 530 nm (2.3 eV), which is responsible for the red colour (Fig. 12(d)). The absorption below 400 nm is associated with interband transitions, and the absorption above 650 nm is due to the free electrons.

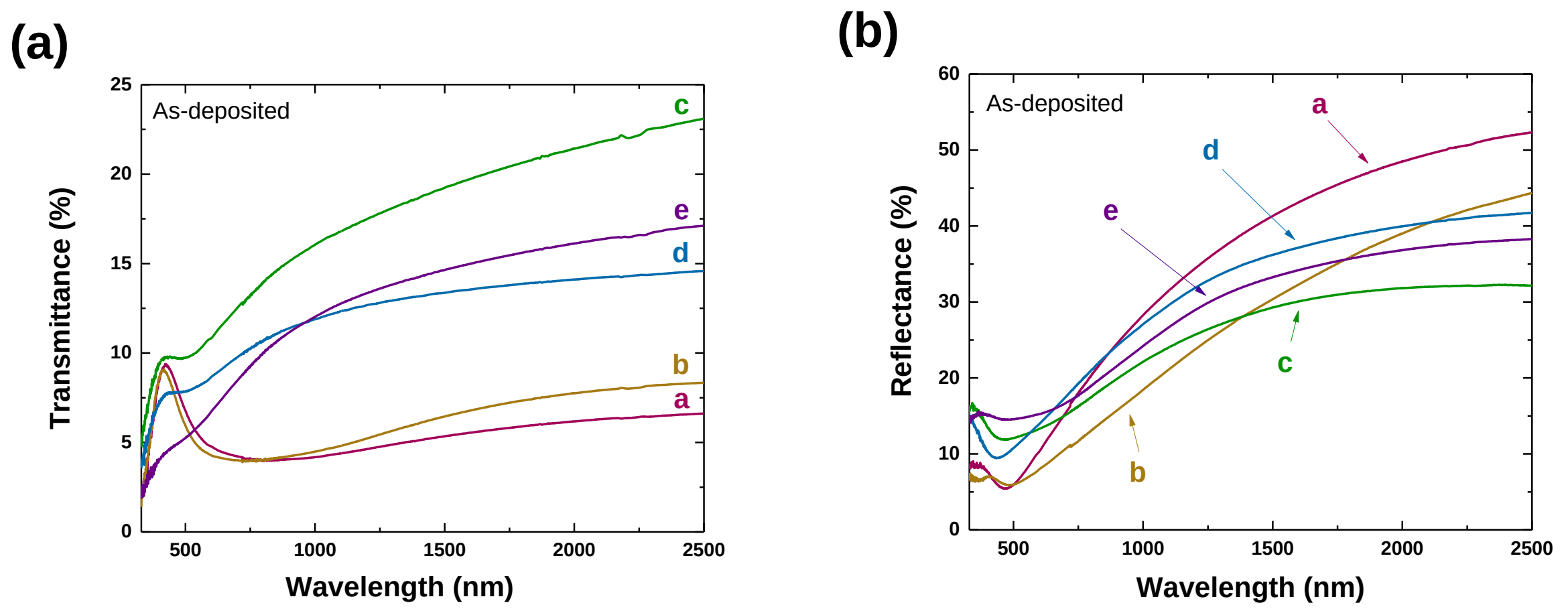

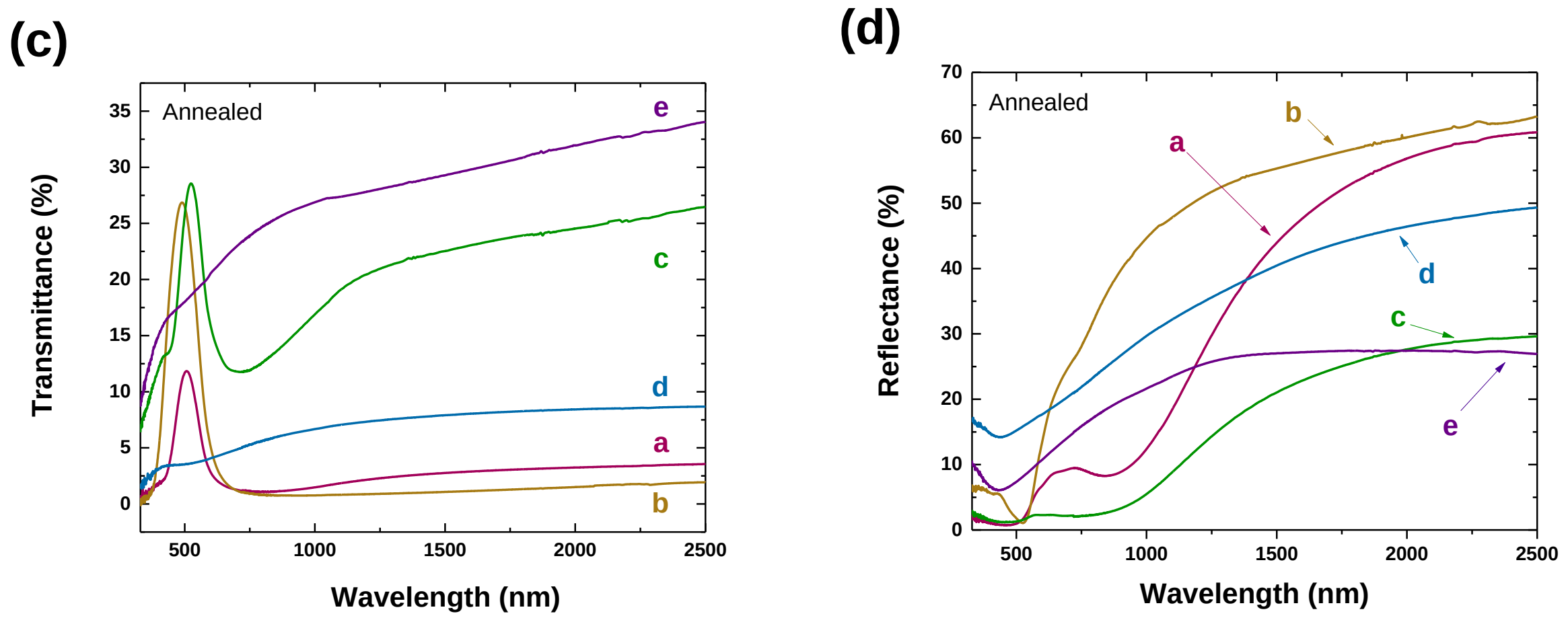

**Fig. 12.** Transmittance (a,c) and specular reflectance (b,d) of the as-deposited and annealed rhenium oxide films, respectively, on fused quartz substrates in the range of 330–2500 nm. The deposition parameters of samples *a*-*e* can be found in Table 1.

Analyzing XRD (Fig. 5 and Fig. 6) and XPS data (Fig. 10 and Fig. 11), we can conclude, that $ReO_x$ film deposited at 250°C consist mainly of the nanocrystalline $ReO_2$ phase. That is why neither XRD nor XPS does not show any changes in the crystal phase or rhenium oxidation state after annealing. It is known, that $ReO_3$ sublimates at a temperature above 400 °C (however, we observed $ReO_3$ sublimation already at 200 °C), and upon heating in a vacuum disproportionate to $Re_2O_7$ and $ReO_2$ [58]. The sublimation temperature of $Re_2O_7$ is 360 °C, while the melting point of $ReO_2$ is above 1000 °C [58]. Optical measurements (Fig. 12) confirm this conclusion: rhenium oxide film as-prepared at 250 °C shows no transparency window around 500 nm (Fig. 12(a) - e), and no transparency window is present also after annealing (Fig. 12(c) - e). A transparency window around 500 nm is associated with the $ReO_3$ phase [14]. The situation is the opposite with RT deposited rhenium oxide thin films. Films as-deposited at RT are X-ray amorphous (showing $Re_2O_7$ crystals only), while after annealing at 250 °C clear $ReO_3$ phase appears (Fig. 5(a,b) and Fig. 6(a,b)). XPS measurements of both as-deposited and annealed RT thin films show the presence of $Re^{6+}$ and $Re^{7+}$ (associated with $ReO_3$ and $Re_2O_7$, respectively). Optical measurements agree well with XRD and XPS data. As-prepared RT thin films have a clearly visible transparency window around 500 nm (Fig.12(a) - a, b), which becomes much more pronounced after annealing (Fig. 12(c) a,b). Thus, by properly combining deposition temperature and post-annealing, it is possible to get either $ReO_3$ or $ReO_2$ thin-film coatings.

## 4. Conclusions

This study reports on the novel approach for the deposition of $ReO_x$ ($x \approx 1.6$–$2.9$) thin films using reactive high power impulse magnetron sputtering (r-HiPIMS) from a metallic rhenium target in a mixed $Ar/O_2$ atmosphere. The influence of the following annealing in air on the properties of rhenium oxide films is also shown. The deposited thin films were characterized by X-ray diffraction, scanning electron microscopy, X-ray photoelectron spectroscopy, electrical resistivity, and optical measurements.

The Re HiPMS discharge current waveforms in a pure Ar atmosphere confirm the gas-sustained self-sputtering regime when operating above the critical current density of 0.15 $A/cm^2$. The peak current decreases significantly when oxygen gas is introduced into the chamber.

The as-deposited at room temperature $ReO_x$ ($x \approx 2.2$) films are X-ray amorphous, and their surface contains a large number of open cavities and voids. When these films are annealed under a capping quartz slide at 250 °C in air, the structural transformation into the nano-crystalline phase of $ReO_3$ occurs. The chemical composition of the annealed film deposited at RT and $O_2/Ar = 0.5$ is $ReO_{2.9}$. When elevated substrate temperatures of 150 or 250 °C are used during the deposition, a relatively dense structure of the nano-crystalline β-$ReO_2$ phase grows on the substrate. The produced films have low electrical resistivity and optical properties typical for thin metallic films.

## Acknowledgments

Financial support was provided by ERDF Project no. 1.1.1.1/ 18/A/073. Institute of Solid State Physics, University of Latvia as the Center of Excellence has received funding from the European Union's Horizon 2020 Framework Programme H2020-WIDESPREAD-01-2016- 2017-TeamingPhase2 under grant agreement No. 739508, project CAMART2.